\documentclass[twocolumn,showpacs,preprintnumbers]{revtex4}
\usepackage{bm,enumerate,dcolumn,tikz,graphicx,color,amsmath,amssymb}
\usepackage{epstopdf}
\usepackage{empheq}
\usepackage{capt-of}
\usepackage{pgfplots}

\usepgfplotslibrary{colormaps,external}
\usetikzlibrary{decorations.markings}
\usetikzlibrary{arrows}

\newcommand{\comment}[1]{}
\newcommand{\bra}{\langle}
\newcommand{\ket}{\rangle}
\newcommand{\eye}{{\rm i}}

\newcommand{\velg}{{\textsc{\tiny VG}}}
\newcommand{\leng}{{\textsc{\tiny LG}}}

\definecolor{lightblue}{rgb}{.8, .8, 1}
\definecolor{myyellow}{RGB}{254,241,24}
\definecolor{myorange}{RGB}{234,125,1}
 
\usepackage{array}
\newcolumntype{C}[1]{>{\centering\let\newline\\\arraybackslash\hspace{0pt}}m{#1}}
\newcolumntype{L}[1]{>{\raggedright\let\newline\\\arraybackslash\hspace{0pt}}m{#1}}
\newcolumntype{R}[1]{>{\raggedleft\let\newline\\\arraybackslash\hspace{0pt}}m{#1}}

\begin{document}

\title{
Dynamic polarizability of Rydberg atoms: 
Applicability of near-free electron approximation, gauge invariance 
and the Dirac sea
} 
\author{Turker Topcu and Andrei Derevianko}
\affiliation{Department of Physics, University of Nevada, Reno, NV 89557, USA}
\date{\today}

\begin{abstract}

Ponderomotive energy shifts experienced by Rydberg atoms in optical fields are known to 
be well approximated by the classical quiver energy of a free electron. We examine such 
energy shifts quantum mechanically and elucidate how they relate to the ponderomotive 
shift of a free electron in off-resonant fields. We derive and evaluate corrections 
to the ponderomotive free electron polarizability in the length and velocity 
(transverse or Coulomb) gauges, 
which agree exactly as mandated by the gauge invariance. We also show how the free 
electron value emerges from the Dirac equation through summation over the 
Dirac sea states. 
We find that the free-electron 
AC Stark shift comes as an expectation value of a term proportional to 
the square of the vector potential in the velocity gauge. 
On the other hand, the same dominant contribution can be obtained to first order via 
a series expansion of the exact energy shift from the second order 
perturbation theory in the length gauge. 
Finally, we numerically examine the validity of the free-electron approximation. The 
correction to the free-electron value becomes smaller with increasing principal 
quantum number, and it is well below a per cent for 60s 
states of Rb and Sr away from the resonances. 

\end{abstract}


\maketitle

Many of the prominent schemes for realizing quantum logic gates for quantum 
computing~\cite{JakCirZol00,UrbJohHen09,GaeMirWil09}, and studying few- and many-body physics 
via quantum simulation experiments using cold Rydberg atoms rely on optically trapped Rydberg 
atoms~\cite{WeiMulLes10,BloDalNas12}. Characterizing optical potentials experienced by 
Rydberg atoms plays an essential role in experimental realization of these 
schemes~\cite{SafWalMol10}. 

The trapping AC Stark shift seen by the Rydberg atom is proportional to its dynamic 
polarizability, which is essentially that of a free electron quivering in the laser field. 
The free electron polarizability $\alpha_{\rm e}(\omega)=-e^2/(m_e \omega^2)$, where $\omega$ is 
the field frequency, can be derived from classical arguments by time averaging 
the electron kinetic energy in the oscillating electric field~\cite{YouAndRai10}. 
It can be also obtained quantum mechanically through the first order perturbation theory, 
with the perturbation being proportional to the intensity~\cite{ZhaRobSaf11}. 

In this paper, we examine the applicability of the free electron approximation for 
Rydberg atoms in length and velocity gauges to gain insight as to how the exact AC Stark 
shift relates to the free electron value in both gauges. We show that $\alpha_{\rm e}(\omega)$ 
emerges from a term in the Hamiltonian proportional to the square of the vector potential 
$(A_{\velg})^2$ in the velocity gauge, whereas it can be recovered to the leading order from 
a series expansion of the exact energy shift from the second order perturbation theory in the 
length gauge. We compute the corrections to the free electron polarizability 
and conclude that the correction to $\alpha_{\rm e}(\omega)$ in Rydberg states is relatively 
small away from resonances. Our discussion elucidates that rather than performing an exact 
sum over states calculation to evaluate the Rydberg state polarizabilities in the length gauge, 
it is simpler to calculate it as an expectation value of $(A_{\velg})^2$. 
Throughout this paper, we assume plane wave electromagnetic fields rather than standing wave 
optical traps. Notice that otherwise the optical trap intensity variation can substantially 
modulate the Rydberg electron polarizability~\cite{TopDer13}. 


The paper is organized as follows: 
In the next section, we start by reviewing scalar and vector potentials of traveling waves 
in the length and velocity gauges. In section II, we derive AC Stark shifts and polarizabilities 
in both gauges within the non-relativistic formalism, and examine how they reduce to the free 
electron value. We then derive the free electron polarizability from the fully-relativistic 
Dirac Hamiltonian in section III, and find that it comes from summation over the negative 
energy states in second order perturbation theory. 
We conclude in section IV by presenting numerical results illustrating the errors made by 
approximating the Rydberg state polarizabilities by the free electron values in the case of 
Rb and Sr atoms. We use atomic units throughout this report unless specifically stated otherwise. 

\section{Scalar and vector potentials}
Electromagnetic fields can be expressed in terms of the scalar and the vector 
potentials $\phi$ and $\mathbf{A}$. While the physical fields are unique, there is a 
certain degree of freedom in the choice of the potentials. 
Specifically, electric and magnetic fields in Gaussian 
units are $\mathbf{F} = -\nabla\phi - (\partial \mathbf{A}/\partial t)/c$ and 
$\mathbf{B} = \nabla\times\mathbf{A}$. A gauge transformation 
\begin{eqnarray}\label{gtran}
\phi &\rightarrow& \phi - \frac{1}{c}\frac{\partial \chi}{\partial t} \;, \\
\mathbf{A} &\rightarrow& \mathbf{A} + \nabla\chi \;, 
\end{eqnarray}
leaves the physical quantities $\mathbf{F}$ and $\mathbf{B}$ unchanged, if $\chi$ satisfy 
the Lorentz condition~\cite{Jackson}. 
The gauging function $\chi$ allows us to transform between different representations of 
$\mathbf{A}$ and $\phi$ in different gauges, although the physical quantities, {\it e.g.} 
observables, are gauge invariant. 

The Hamiltonian for an optical electron in external electromagnetic 
field may be written as 
\begin{equation}
H = \frac{1}{2}\left ( \mathbf{p} - \mathbf{A}/c \right )^2 + V_{C} + \phi \;,
\end{equation}
where $V_{C}$ is the core potential seen by the electron. 

In our current discussion, we treat the electromagnetic field as a plane wave propagating along 
the $z$-direction, and describe the electric field as 
$\mathbf{F} = F_0 \;\hat{\epsilon}\; e^{\eye(kz-\omega t)} +{\rm c.c.}$. 
In the velocity gauge (also known as the transverse or Coulomb gauge), 
\begin{eqnarray}
\mathbf{A}_{\velg} &=& -\eye\frac{F_0 c}{\omega} \;\hat{\epsilon}\; 
	e^{\eye(kz-\omega t)} +{\rm c.c.} \;, \nonumber \\
\phi_{\leng} &=& 0 \;. \label{eq:A_vg}
\end{eqnarray}
The length and the velocity gauges are related through the transformation function
$\chi = -\mathbf{r}\cdot\mathbf{A}_{\velg}$, leading to potentials in the length 
gauge~\cite{Jackson}: 
\begin{eqnarray}
\mathbf{A}_{\leng}	&=& -F_0 \;\hat{\mathbf{k}}\;(\hat{\epsilon}\cdot\mathbf{r})
	\; e^{\eye(kz-\omega t)} +{\rm c.c.} \label{eq:A_lg} \;, \nonumber \\
\phi_{\leng} &=& -F_0 \;(\hat{\epsilon}\cdot\mathbf{r})\; 
	e^{\eye(kz-\omega t)} +{\rm c.c.} \label{eq:phi_lg}.
\end{eqnarray}

\section{Stark shifts and polarizabilities in the non-relativistic approximation}
\subsection{Velocity Gauge}
In the velocity gauge~\eqref{eq:A_vg}, the Hamiltonian becomes, 
\begin{equation}\label{eq:hamiltonian}
H = \frac{p^2}{2} +V_{C} - \frac{\mathbf{A}_{\velg}\cdot\mathbf{p}}{c} 
	+ \frac{(A_{\velg})^2}{2c^2} .
\end{equation}
We calculate the energy shifts within the Floquet formalism of quasi-energy states~\cite{ManOvsRap86} 
for the third term, and use the first order perturbation theory for the fourth term since 
it is already second order in the field strength. 
The energy shift for a Rydberg state $|r\ket$ therefore becomes 
\begin{align}
&\begin{aligned}
	\delta E_{\velg}(\omega) = \frac{1}{4c^2} 
		&\sum_j \frac{2\Delta E_j}{\Delta E_j^2-\omega^2} 
			|\bra r|\mathbf{A}_{\velg}\cdot\mathbf{p}|j\ket|^2 \nonumber \\
			&+ \left\bra r \left|({A}_{\velg})^2 \right|r \right\ket/(2c^2) \;. \nonumber 
	\end{aligned}
\end{align}
The first term can be decomposed into the scalar, vector and tensor contributions. 
We focus on the dominant scalar contribution to the Stark shift and arrive at 
\begin{align}
\delta E_{\velg}(\omega) = \frac{F_0^2}{4\omega^2} 
	+ \frac{F_0^2}{4\omega^2} \sum_j 
	\frac{2\Delta E_j^3}{\Delta E_j^2-\omega^2} 
	\frac{1}{3} |\bra r|\mathbf{D}|j \ket|^2  \;, \label{eq:e2nd_vg}
\end{align}
\comment{
\begin{align}
&\begin{aligned}
	\delta E_{\velg}&(\omega) = \frac{1}{4c^2} \Bigg (
		\sum_j \frac{\bra r|\mathbf{A}_{\velg}\cdot\mathbf{p}|j\ket 
			\bra j|\mathbf{A}_{\velg}\cdot\mathbf{p}|r\ket}{\Delta E_j +\omega^2} \nonumber \\
		&+\sum_j \frac{\bra j|\mathbf{A}_{\velg}\cdot\mathbf{p}|r\ket 
			\bra r|\mathbf{A}_{\velg}\cdot\mathbf{p}|j\ket}{\Delta E_j -\omega^2} \Bigg )
		+ \left\bra r \left|\frac{{A}_{\velg}^2}{2c^2} \right|r \right\ket \nonumber 
	\end{aligned}\\ 
&\delta E_{\velg}(\omega) = \frac{F_0^2}{4\omega^2} \sum_j 
	\frac{2\Delta E_j^3}{\Delta E_j^2-\omega^2} 
	\frac{1}{3} |\bra r|\mathbf{D}|j \ket|^2  
	+ \frac{F_0^2}{4\omega^2} \label{eq:e2nd_vg}
\end{align}
}
where $E_r-E_j = \Delta E_j$, and we used 
\begin{eqnarray}
\bra r|\hat{\epsilon}\cdot\mathbf{p}|j\ket = -\eye \;\Delta E_j\;\hat{\epsilon}\; 
	\bra r|\mathbf{D}|j\ket. 
\end{eqnarray}
Here $\mathbf{D}$ is the usual dipole operator. The summation over the magnetic quantum 
number can be evaluated explicitly using the Wigner-Eckart theorem: 
\begin{eqnarray}\nonumber
\sum_{\lambda,M_j} (-1)^{\lambda} \bra r|\mathbf{D}_\lambda|j\ket 
		\bra j|\mathbf{D}_{-\lambda}|r\ket 
			= \frac{1}{2l_r +1} \sum_{n_j,l_j} |\bra r||D||j\ket|^2 \;, 
\end{eqnarray}
where $\bra r||D||j \ket$ are the reduced dipole matrix elements. 

The first term in Eq.~\eqref{eq:e2nd_vg} can be immediately identified as the ponderomotive 
energy shift for a free electron quivering in electromagnetic field. Notice that it came 
directly from the $A_{\velg}^2$ term in the Hamiltonian~\eqref{eq:hamiltonian}, and it 
is the dominant term for a Rydberg state. By contrast, the second term in~\eqref{eq:e2nd_vg} 
is the correction to the free electron ponderomotive shift. 
All of the resonance structure is included in the second correction term, whereas 
the free electron shift only provides a smooth background on which the resonance structure 
sits. 

We can express $\delta E_{\velg}$ 
in terms of the conventional AC polarizability through its definition for the plane waves, 
$\delta E(\omega) = -\alpha(\omega) F_0^2/4$. Then the AC polarizability 
in the velocity gauge becomes 
\begin{eqnarray}
\alpha_{\velg}(\omega) &=& \alpha_{\rm e}(\omega) 
	-\frac{2}{3\omega^2} \sum_j 
	\frac{\Delta E_j^3}{\Delta E_j^2-\omega^2} 
	|\bra r|\mathbf{D}|j \ket|^2  \;. \label{eq:alp_vg_exact} 
\end{eqnarray}
Here $\alpha_{\rm e}(\omega) = - 1/\omega^2$ is the free electron polarizability. 

\subsection{Length Gauge}
We now derive Eq.~\eqref{eq:alp_vg_exact} in the length gauge, and thereby establish the 
gauge invariance. Using the length gauge vector and 
scalar potentials~\eqref{eq:phi_lg}, the Hamiltonian can be expressed as 
\begin{equation}
H = \frac{p^2}{2} + V_{C} - \frac{\mathbf{A}_{\leng}\cdot\mathbf{p}}{c} 
	+ \frac{(A_{\leng})^2}{2c^2} +\phi_{\leng} .
\end{equation}
To calculate the energy shifts resulting from the last three terms in the Hamiltonian, 
we again use the quasi-energy formalism~\cite{ManOvsRap86} for the third and the fifth 
terms, and the first order perturbation theory for the fourth term: 
\begin{align}
\delta E_{\leng}(\omega) = \frac{F_0^2}{4} &\sum_j 
	\frac{2\Delta E_j /3}{\Delta E_j^2-\omega^2} 
	|\bra r|\mathbf{A}_{\leng}\cdot\mathbf{p}/c + \phi_{\leng} |j \ket|^2 \nonumber \\
	&+ \frac{F_0^2}{2c^2} \bra r|(A_{\leng})^2|r\ket  \;. 
\end{align}
Upon expanding the square, we encounter three contributions to the first term. 
Out of the three, the dominant one is the $|\bra r|\phi_{\leng}|j \ket|^2$ term. The 
term involving $|\bra r|\mathbf{A}_{\leng}\cdot\mathbf{p}|j \ket|^2$, and the last 
term involving $(A_{\leng})^2$ are suppressed by a factor of $1/c^2 \sim 10^{-4}$ and 
can be safely ignored in the non-relativistic approximation. 
The cross term between the $\mathbf{A}_{\leng}\cdot\mathbf{p}$ 
and $\phi_{\leng}$ drops out because it is proportional to 
${\rm Re} [ \bra r|\mathbf{A}_{\leng}\cdot\mathbf{p}|j \ket \bra j|(\phi_{\leng})^\dagger|r \ket ]^2$ 
which vanishes on the account of 
$\bra r|\mathbf{A}_{\leng}\cdot\mathbf{p}|j \ket$ being purely imaginary (due to 
$\bra r|\hat{\epsilon}\cdot\mathbf{p}|j\ket = -\eye\Delta E_j\;\hat{\epsilon}\;\bra r|\mathbf{D}|j\ket$). 
We again focus on the dominant scalar contribution to the Stark shift, which leaves us 
with
\begin{align}
\delta E_{\leng}(\omega) = \frac{F_0^2}{2} \frac{1}{3} \sum_j 
	\frac{\Delta E_j}{\Delta E_j^2-\omega^2} 
	|\bra r|\mathbf{D}|j \ket|^2 .
\end{align}
Through the definition $\delta E(\omega) = -\alpha(\omega) F_0^2/4$, the polarizability in the 
length gauge can be written as:
\begin{align}
\alpha_{\leng}(\omega) = -\frac{2}{3} \sum_j 
	\frac{\Delta E_j}{\Delta E_j^2-\omega^2} 
	|\bra r|\mathbf{D}|j \ket|^2 . \label{eq:alp_lg_exact}
\end{align}
To reveal how this expression relates to $\alpha_{\rm e}(\omega)$ and the polarizability in 
the velocity gauge, 
we expand the resolvent operator in Eq.~\eqref{eq:alp_lg_exact} in powers of $\Delta E_j/\omega$ :
\begin{align}\label{eq:expansion}
\frac{\Delta E_j}{\Delta E_j^2-\omega^2} = \frac{\Delta E_j}{\omega^2} 
		\left(-1-\frac{\Delta E_j^2}{\omega^2}-\frac{\Delta E_j^4}{\omega^4}-\cdots  \right) .
\end{align}
This results in a series expansion of the polarizability in the length gauge:
\begin{align} 
{\alpha}_{\leng}(\omega) = &\frac{2}{3\omega^2} \sum_j 
	\Delta E_j \;|\bra r|\mathbf{D}|j \ket|^2 \nonumber \\
		&+ \frac{2}{3\omega^4} \sum_j \Delta E_j^3 \;|\bra r|\mathbf{D}|j \ket|^2 \nonumber \\
		&+ \frac{2}{3\omega^6} \sum_j \Delta E_j^5 \;|\bra r|\mathbf{D}|j \ket|^2 
		+ \cdots \;. \nonumber  
\end{align}
This series can be resummed such that the first term is separated out, 
\begin{align} \label{eq:alp_lg_expanded}
{\alpha}_{\leng}(\omega) = &\frac{2}{3\omega^2} \sum_j 
	\Delta E_j \;|\bra r|\mathbf{D}|j \ket|^2 \nonumber \\
	&+ \frac{2}{3\omega^2} \sum_k \left( 
			\sum_j \frac{\Delta E_j^{3+2k}}{\omega^{2+2k}} \;|\bra r|\mathbf{D}|j \ket|^2 
		\right ) \;. 
\end{align}
With the help of the oscillator sum rule, 
\begin{align}
-\frac{2}{3} \sum_j \Delta E_j \;|\bra r|\mathbf{D}|j \ket|^2 = 1 \;, 
\end{align}
we recover the free electron polarizability from the first term in~\eqref{eq:alp_lg_expanded} 
when summed over a complete set of states (note that $\Delta E_j = -(E_j - E_r)$). 
On the other hand, with the aid of the expansion~\eqref{eq:expansion}, the outer sum in the 
second term can be collapsed back into the same correction term to the free 
electron polarizability in the velocity gauge~\eqref{eq:alp_vg_exact}. Thus 
\begin{align} 
{\alpha}_{\leng}(\omega) = \alpha_{\rm e}(\omega) 
	+ \frac{2}{3\omega^4} \sum_j 
		\frac{\Delta E_j^3}{\Delta E_j^2-\omega^2}  
			\;|\bra r|\mathbf{D}|j \ket|^2 \;, \label{eq:alp_lg_exact2}
\end{align}
and ${\alpha}_{\leng}(\omega) \equiv {\alpha}_{\velg}(\omega)$. 
This expression for the length gauge polarizability is identical to the one 
in Eq.~\eqref{eq:alp_vg_exact} in the velocity gauge, confirming the equivalence of 
the AC Stark shifts in both gauges, {\it i.e.} the gauge invariance. These expressions also show 
how the free electron term originates, and we will demonstrate below that the correction to 
$\alpha_{e}(\omega)$ is indeed very small for Rydberg states. 

In Ref.~\cite{SafWilCla03}, the emergence of the free electron polarizability in the length 
gauge has been shown numerically for Rydberg states of Rb atom. Ref.~\cite{Ovsiannikov80} has 
provided an alternative analytical expression using higher rank oscillator sum rules. However, 
we find that the formula of Ref.~\cite{Ovsiannikov80} has poor convergence properties when 
evaluated numerically. 

\section{Dirac Sea}
\label{Sec:DiracSea}
So far we examined the AC Stark effect in the non-relativistic formalism. 
In the velocity gauge the dominant contribution came from the expectation value of the 
vector-potential-squared term $A_{\velg}^2$. Curiously, the fully-relativistic Dirac Hamiltonian 
involves only linear couplings to electromagnetic fields, thereby such term is missing in the 
lowest-order of  perturbation theory. This naturally raises a question of just how the 
free-electron polarizability emerges from the relativistic equations. As we show in this Section, 
it comes from summations over the negative energy (positron, $E_n < -m_e c^2$)  states, i.e. 
from the Dirac sea. Similarly pronounced effects of Dirac sea on weak atomic transition amplitudes 
were found earlier~\cite{DerSavJoh98,SavDerBer99} in relativistic many-body calculations. 

We start with the coupling to electromagnetic fields 
\begin{equation}
V = \mathbf{\alpha}\cdot\mathbf{A}(\mathbf{r}, t) - \phi(\mathbf{r}, t) \, ,
\end{equation}
where $\mathbf{\alpha}$ are the conventional Dirac matrices.  In the velocity gauge
$\phi_{\velg}(\mathbf{r}, t)=0$, so that
\begin{equation}
V = \mathbf{\alpha}\cdot\mathbf{A}_{\velg}(\mathbf{r}, t) .
\end{equation}
Applying the second-order Floquet formalism, we arrive at the fully-relativistic AC polarizability 
\begin{equation}\label{dynpol}
\begin{split}
\alpha_{\velg,\mathrm{Dirac}}(\omega)  = -\frac{2c^2}{\omega^2}
  \sum_{j}&\;\frac{(E_r-E_j)}{(E_r-E_j)^2-\omega^2} \\
  &\;\;\times|\bra\psi_r|\mathbf{\alpha}\cdot\hat{\epsilon}\;
    |\psi_j\ket|^2 \, .
\end{split} 
\end{equation} 

The summation over intermediate states in the above equation spans the complete spectrum of the 
Dirac equation, i.e., both positive and negative energy states. It can be shown using the Pauli 
approximation that the sum over the conventional positive energy states recovers the second term 
(second-order sum) in the non-relativistic expression for the AC polarizability.  Now we demonstrate 
that the dominant, free-electron polarizability term, emerges from summing over the negative  energy 
states. In this case, the Rydberg electron energy $E_r$ is above the Dirac sea level and  
$(E_r-E_j)\approx 2m_ec^2$ and $(E_r-E_j)\gg \omega$ (we naturally assume that the photon energy 
is well below $2m_e c^2$). Within this approximation, the dynamic AC polarizability becomes
\begin{align}\label{dynpol_exp}
\alpha_{\velg,\mathrm{Dirac}}^{(-)} (\omega)   = -&\frac{2c^2}{\omega^2(2m_ec^2)} \nonumber \\ 
	&\times \sum_{E_j < -m_ec^2} 
	  \bra\psi_r|\mathbf{\alpha}\cdot\hat{\epsilon}\;|\psi_j\ket 
	  \bra\psi_j|\mathbf{\alpha}\cdot\hat{\epsilon}\;|\psi_r\ket \, .
\end{align} 
The Dirac matrices $\alpha$  mix the large and the small components of the Dirac bi-spinor, 
$\psi =
 \begin{pmatrix}
 \psi^{l} \\
 \psi^{s} 
\end{pmatrix}$. 
Part of the summation on the right hand side of~\eqref{dynpol_exp} can therefore be written as 
\begin{equation}
\sum_{E_j < -m_ec^2 } 
  \bra\psi^l_r|\mathbf{\sigma}\cdot\hat{\epsilon}\;
    |\psi^s_j\ket 
  \bra\psi^s_j|\mathbf{\sigma}\cdot\hat{\epsilon}\;
    |\psi^l_r\ket \, .
    \end{equation} 
In the non-relativistic approximation, the negative energy states are complete amongst 
themselves, i.e. $\sum_b |\psi^s_b\ket\bra\psi^s_b|\approx 1$, leading to
\begin{equation}
\alpha_{\velg,\mathrm{Dirac}}^{(-)} (\omega)    = -\frac{1}{\omega^2} 
  \bra\psi^l_a| \mathbf{\sigma}\cdot\hat{\epsilon}\;
    |\psi^l_a\ket =  -\frac{1}{\omega^2} \, 
\end{equation}
where we used $(\mathbf{\sigma}\cdot\hat{\epsilon})^2 = 1$ and $m_e=1$ a.u..
Therefore  we recover  the free electron polarizability in atomic units. 
Is not interesting that  describing purely classical effect of electron quiver motion 
requires the notion of anti-mater? 


\section{Correction to the free electron polarizability}
In all our calculations below, we use the 
single active electron approximation, and represent the atomic core with the model 
potential using the correct quantum defects for Rb and Sr atoms from~\cite{AymGreLuc96}: 
\begin{equation} \label{eq:quant_defect_pot}
V_{C} = -\frac{1+2e^{-\alpha_1r}+r\alpha_2 e^{-\alpha_3r}}{r}
-\frac{\alpha_s}{2r^4}[1-\exp(-r^3)]^2 \;, 
\end{equation}
where $\alpha_1$, $\alpha_2$ and $\alpha_3$ are $l$-dependent screening parameters, and 
$\alpha_s$ is the static polarizability of the core~\cite{SahTimDas09}. For illustrative purposes, 
we pick Rb and Sr atoms because they have very different 
core polarizabilities. For Rb, the static core polarizability is $\sim$9 a.u., whereas 
the polarizability for the Sr$^+$ ion with a 5s valence electron serving as spectator is $\sim$91 a.u., 
an order of magnitude larger. This results in large differences in quantum defects $\delta_l$ 
for the $s$ Rydberg states: for example in the 100s state of Rb $\delta_0 \simeq 3.28$ 
and for Sr $\delta_0 \simeq 5.02$. 
In our calculations, we assume that the Sr atom is in 
a $J=0$ state, and we ignore contributions from the resonant structure from the 
5s valence electron. 

In order to evaluate polarizabilities in both the length and the velocity gauges, 
we have to evaluate sums over a complete set of states in~\eqref{eq:alp_vg_exact} 
and~\eqref{eq:alp_lg_exact}. We check the completeness of our basis using the 
well known oscillator sum rule. We find, 
\begin{align}
-\frac{2}{3} \sum_j \Delta E_j \;|\bra r|\mathbf{D}|j \ket|^2 &= 0.967 
	\;\;\;{\rm (Rb)} \label{eq:sum_rule_rb} \;, \\
-\frac{2}{3} \sum_j \Delta E_j \;|\bra r|\mathbf{D}|j \ket|^2 &= 0.989  
	\;\;\;{\rm (Sr)} \label{eq:sum_rule_sr} \;,
\end{align}
while summing over $p$ states with $n=2$ through 200 for the 100s states of the Rb and Sr atoms. 

\begin{figure}[h!tb]
	\begin{center}$
		\begin{array}{c}
                \resizebox{85mm}{!}{\includegraphics{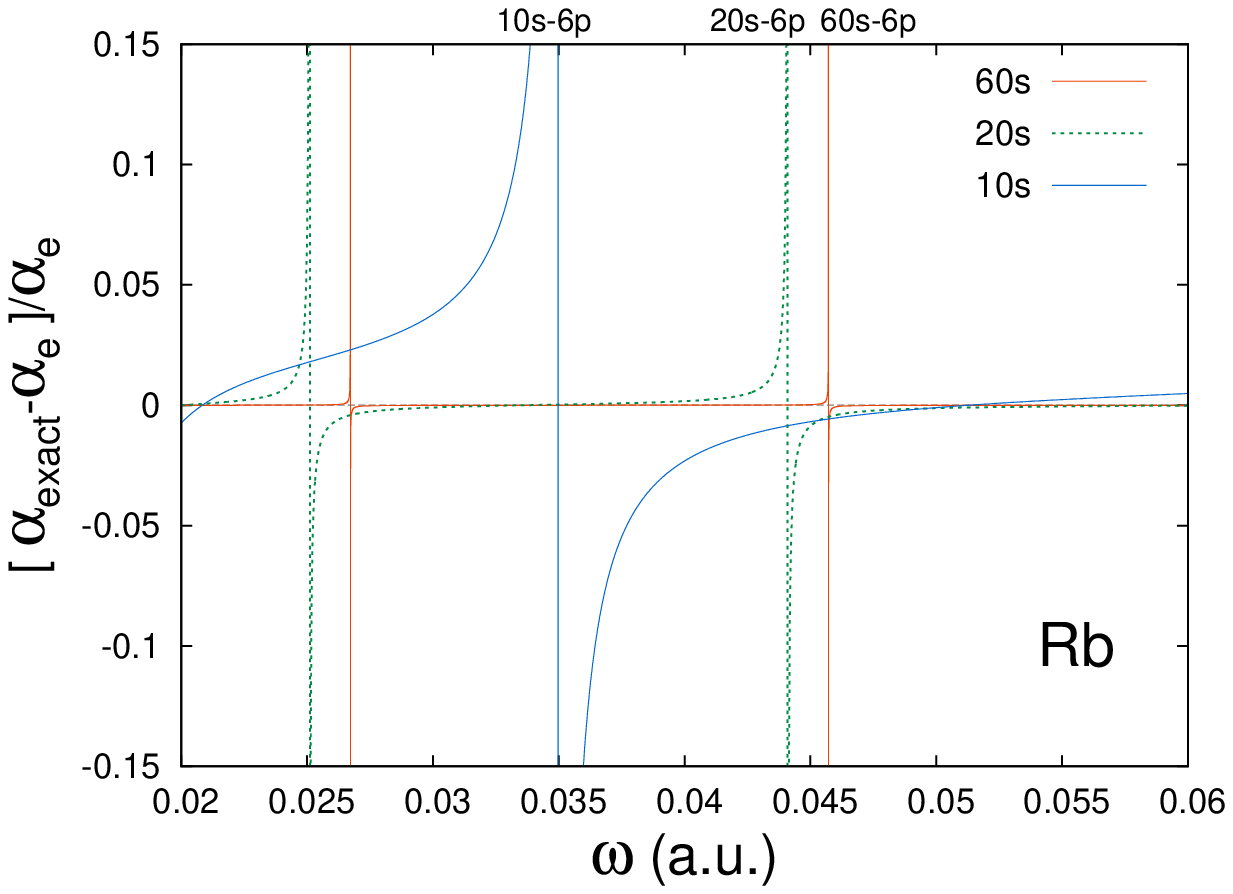}} \\ 
                \resizebox{85mm}{!}{\includegraphics{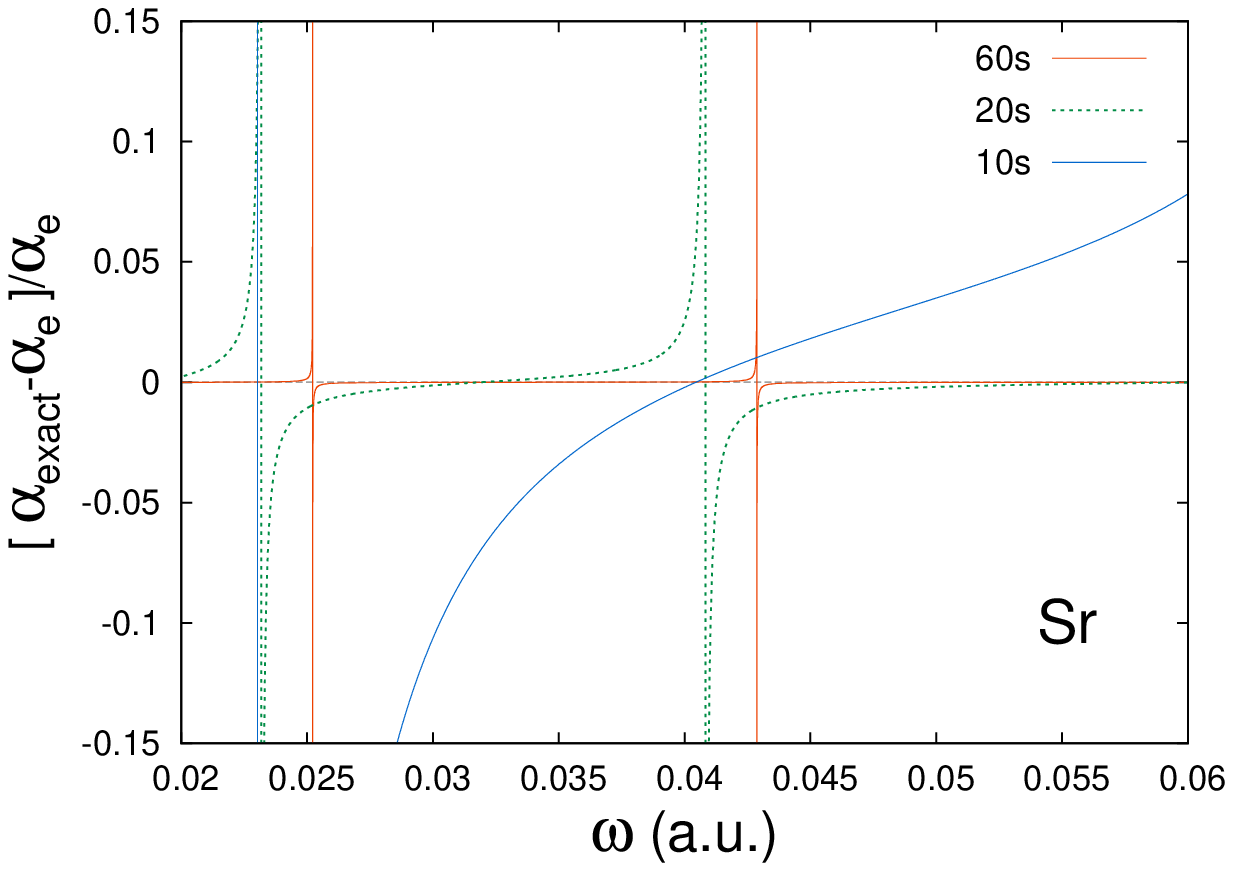}} 
         \end{array}$
	\end{center}
	\caption{(Color Online) The fractional corrections to the free electron polarizability 
	for the 60s, 20s and 10s states of Rb and Sr atoms in the IR region of the spectrum. 
	The resonance structure is entirely contained within these corrections, and the resonances 
	become wider for higher $n$s states. The $\omega$ axis spans wavelengths between 
	$\sim$2300 nm and $\sim$760 nm. 
	}
	\label{fig:01}
\end{figure}

We have shown that the Rydberg state polarizability in the length~\eqref{eq:alp_lg_exact2} 
and velocity gauges~\eqref{eq:alp_vg_exact} are identical as a result of the gauge invariance, 
{\it i.e.} $\alpha_{\leng}(\omega)\equiv\alpha_{\velg}(\omega)$. Furthermore, we have expressed 
these exact polarizabilities as a sum of the free electron polarizability $\alpha_{\rm e}(\omega)$  
and an exact correction. The resonance structure of the polarizabilities is entirely 
captured by this correction term, whereas $\alpha_{\rm e}(\omega)$ only provides a smooth 
background. The fractional corrections for various $n$s states of Rb and Sr are shown in 
Fig.~\ref{fig:01} in the IR region of the spectrum. The resonant structure is evident in the 
plots, and the widths of the resonances increase with increasing principal quantum number. Also, 
the sizes of the corrections grow smaller as $n$ is increased meaning that the near-free electron 
approximation to the Rydberg electron polarizability gradually becomes more accurate away from the 
resonances. With the exception of the 10s state, the corrections are well below a per cent for 
both Rb and Sr, except at the resonances. 

Comparing the upper and lower panels of Fig.~\ref{fig:01}, we observe that 
the errors made by approximating $\alpha_{\leng}(\omega)$ and $\alpha_{\velg}(\omega)$ 
by $\alpha_{\rm e}(\omega)$ 
are larger for Sr than for Rb. This stems from the larger static polarizability 
of the Sr$^+$ ion with a 5s valence electron compared to the Rb$^+$ ionic core, a property 
which is contained in the model potential~\eqref{eq:quant_defect_pot}. Although 
the static core polarizabilities for these atoms differ by an order of magnitude, 
the corrections to $\alpha_{\rm e}(\omega)$ are still much less than a percent in 
both cases, except in the immediate vicinity of resonances. 

Another feature seen in Fig.~\ref{fig:01} is that the widths of the resonances decrease 
with increasing principal quantum number. For example, for Rb the width of the 20s-6p 
resonance is significantly larger than that of the 60s-6p resonance. On the other hand, the 
10s-6p resonance is so large that the correction to $\alpha_{\rm e}(\omega)$ for the 10s 
state of Rb is at least a couple of per cent nearly for all frequencies spanned in 
Fig.~\ref{fig:01}. Similar observations can also be made for Sr. 
The widening of these resonances with increasing $n$ can be qualitatively understood if 
one realizes that the widths of the resonances in Eqs.~\eqref{eq:alp_vg_exact} 
and~\eqref{eq:alp_lg_exact2} are controlled by the square of the 
dipole matrix elements, which scales as $|\bra r|D|j\ket|^2 \sim 1/n^3$. 

Ref.~\cite{TopDer13} found strong intensity landscape modulations of the effective 
polarizability for Rydberg atoms trapped in IR lattices. Because of the low frequency 
of the trapping field, the Rydberg electron polarizability can potentially deviate 
quite substantially from the free electron value. This is because at sufficiently low frequencies 
the polarizability must approach its static limit, whereas the free electron value diverges. 
In this paper, we have shown that the free electron approximation holds, even at IR 
wavelengths of thousands of nm. 

This work was supported by the NSF Grant No. PHY-1212482.

\end{document}